\documentclass[preprint2]{aastex6}

\usepackage{amsmath}
\usepackage{units}

\shorttitle{Integral upper limits assuming power law spectra}
\shortauthors{Ahnen}
\begin{document}


\title{On integral upper limits assuming power law spectra \\and 
the sensitivity in high-energy astronomy}

%% Use \author, \affil, plus the \and command to format author and affiliation 
%% information.  If done correctly the peer review system will be able to
%% automatically put the author and affiliation information from the manuscript
%% and save the corresponding author the trouble of entering it by hand.
%%
%% The \affil should be used to document primary affiliations and the
%% \altaffil should be used for secondary affiliations, titles, or email.

%% Authors with the same affiliation can be grouped in a single
\author{Max L. Ahnen\altaffilmark{1}}
\affil{Institute for Particle Physics, ETH Zurich, 8093 Zurich,
Switzerland}


\altaffiltext{1}{m.knoetig@gmail.com}

\begin{abstract}

The high-energy non-thermal universe is dominated by power law-like
spectra. Therefore results in 
high-energy astronomy are often reported 
as parameters of power law fits, or, in the case of a non-detection, as an
upper limit assuming the underlying unseen spectrum behaves as a power law.
In this paper I demonstrate a simple and powerful
one-to-one relation of the integral upper limit in the two dimensional
power law parameter space into
the spectrum parameter space and use this method to
unravel the so far convoluted question of the sensitivity of
astroparticle telescopes.

\end{abstract}




\section{Introduction} \label{sec:intro}
Power law spectra are a general feature in the 
non-thermal high-energy universe. 
Often results in high-energy astronomy are reported as parameters 
of power law fits~\citep[e.g.][]{Ackermann2016FHL,Aharonian2006GalCenter}, 
or, if no source is detected, as a 
flux limit~\citep[e.g.][]{Archambault2016BlazarsUL,Aliu2012Nova,Aliu2014GRB_UL,
Ahnen2016Geminga,Aleksic2015BinaryUL,Abramowski2014AGN_UL,
Aliu2015VeritasGeminga,Ahnen2016Perseus} assuming 
the unseen spectrum $dN/dE$ is a power law
\begin{equation}
\frac{dN}{dE} = f_0 \left(\frac{E}{E_0}\right)^{\Gamma} .
\label{eqn:power_law}
\end{equation}
In Eqn.~\ref{eqn:power_law} $E_0$ is a freely chosen reference energy, 
$f_0$ is the flux normalization, and $\Gamma$ is the spectral index.
 
The ability of gamma-ray, neutrino, and cosmic-ray telescopes to detect a source 
depends on their intrinsic parameters and on the external source spectrum
~\citep{Aleksic2016Performance,Aharonian2006Performance,Murase2008,Bertou2002,Aartsen2015}.
However, even when no significant signal counts are
detected, lone would like to put 
limits on the observed signal flux 
would allow inference of physical results, 
as observing nothing is fundamentally different
from not observing at all.
But what does an upper limit on the signal count rate
mean for the spectrum? This question is tightly related to
the detection limit 
of a given instrument. After all, every observer is 
interested to know: How long does it take to detect my source
with an assumed average flux?

Currently, there are only 
convoluted or partial answers to these two fundamental questions 
in high-energy astronomy.
However, the universal observation of power laws 
in gamma-ray sources allows framing the questions in the 
$\{f_0,\Gamma\}$ parameter space of the power law. 

\section{Upper limits in the context of On/Off measurements} \label{sec:context}
In the low count regime of high-energy astronomy one experimental 
method is well established: 
The On/Off measurement~\citep{Li1983,Rolke2005,Knoetig2014,Abraham2007,Abbasi2011}.
In it, the 
experimenter would like to measure an imprecisely known 
background count rate as well as a potential signal count rate. 
The measurement consists of two observations:
The first counts $N_\text{on}$ 
events in a region with potential signal,
the second counts $N_\text{off}$ events in an adjacent 
region which is assumed to be signal free. Both regions 
are related to one another through the ratio of exposures
$\alpha$. If the data show no evidence for a source, 
an upper limit on the signal counts 
$\lambda_\text{lim}$ can be calculated
using $N_\text{on}$, $N_\text{off}$, and $\alpha$~
\citep{Rolke2005,Knoetig2014}.
This limit can be used to infer constraints on the potential source spectrum. 

The most constraining upper limits can be extracted from 
integral upper limits, calculated using
the average signal counts $\lambda_\text{s}$ integrated over 
all energies. Then, in order to 
to make inferences about the source spectrum one, usually, further assumes
a certain power law with a common 
spectral index $\Gamma$
~\citep{Abramowski2014AGN_UL,Aleksic2015BinaryUL,Ahnen2016Geminga,
Aliu2015VeritasGeminga,Aliu2014GRB_UL}. 
\begin{eqnarray}
\lambda_\text{s} &=& t \int \, \frac{dN}{dE} \, A_\text{eff}(E) \, dE , \\
 &=& t \, f_0 \int \, \left(\frac{E}{E_0}\right)^{\Gamma}  \, A_\text{eff}(E) \, dE .
\label{eqn:signal_counts}
\end{eqnarray} 
Assuming a steady source, the integrated average 
signal counts $\lambda_\text{s}$ depend
on the instrument effective area for each energy $A_\text{eff}(E)$, 
the observation time 
$t$, and the assumed source spectrum parameter $\Gamma$.
The limit flux normalization 
$f_0$ 
is then calculated such that the resulting power law spectrum 
would yield the limit 
%$\lambda_\text{s} = \lambda_{lim}$.
\begin{equation}
\lambda_\text{s} = \lambda_\text{lim}.
\label{eqn:exclusion_curve}
\end{equation}
The result
is often reported as integrated source flux $F_{>E_\text{thr}}$
above a certain threshold energy $E_\text{thr}$
\begin{equation}
F_{>E_\text{thr}} = \int^{\infty}_{E_\text{thr}} \, \frac{dN}{dE} \, dE.
\label{eqn:integral_flux}
\end{equation}

Implicit in this method is the choice of the power law index $\Gamma$
which can be extrapolated~\citep{Ahnen2016Geminga}, physically motivated
~\citep{Aliu2015VeritasGeminga,Aliu2014GRB_UL},
or hedged to have little impact on the result~\citep{Abramowski2014AGN_UL,
Aleksic2015BinaryUL}. However, extrapolating is not always possible and
the power law index often varies strongly from source to 
source~\citep{Ackermann2016FHL}, which propagates to 
a potentially large systematic error in the flux limits.

A method to reduce the systematic error of the choice of
power law index and to infer spectral information in the absence of signal
is to calculate differential upper limits.
Although they are called differential, no observation can 
measure differentials directly. Therefore,
differential upper limits are approximated by upper 
limits integrated in small bins, assuming a certain power law index 
$\Gamma$ and using Eqn.~\ref{eqn:signal_counts} and Eqn.~\ref{eqn:exclusion_curve}.
The limit curve obtained in 
this way still depends on the choice of $\Gamma$~\citep{Ahnen2016Geminga},
admittedly less than the full integral upper limit. The main 
issue with differential upper limits is their lower sensitivity due to 
the binning, which 
falls short of the real detection capabilities of the instrument.

A more sensitive approach to the influence of the power law index $\Gamma$
is the decorrelation energy method~\citep{Aliu2012Nova,
Archambault2016BlazarsUL}. Here, integral upper limits are first calculated 
assuming three different power law indices. This results in three 
different limit power laws. 
The decorrelation energy is then defined as the energy at which their flux 
and therefore their upper limit estimate depends the least 
on the power law index.
In a final step the weakest of the three upper limits at the decorrelation 
energy (which turns out to be the 
upper limit of the central power law index) is reported. 
While this method better accounts for variations in $\Gamma$,
it is still not universal as three common power law indices have to be
chosen, none of which are distinguished. 

In this paper, I demonstrate a more useful
approach to integral upper limits
than the previously discussed methods. The key insight comes from 
investigating integral upper limits
in the two dimensional $\{f_0,\Gamma\}$ 
parameter space of the power law.

\section{Generalized integral upper limits assuming power law spectra} \label{sec:method}
\begin{figure*}
\centering
\includegraphics[width=0.49\textwidth]{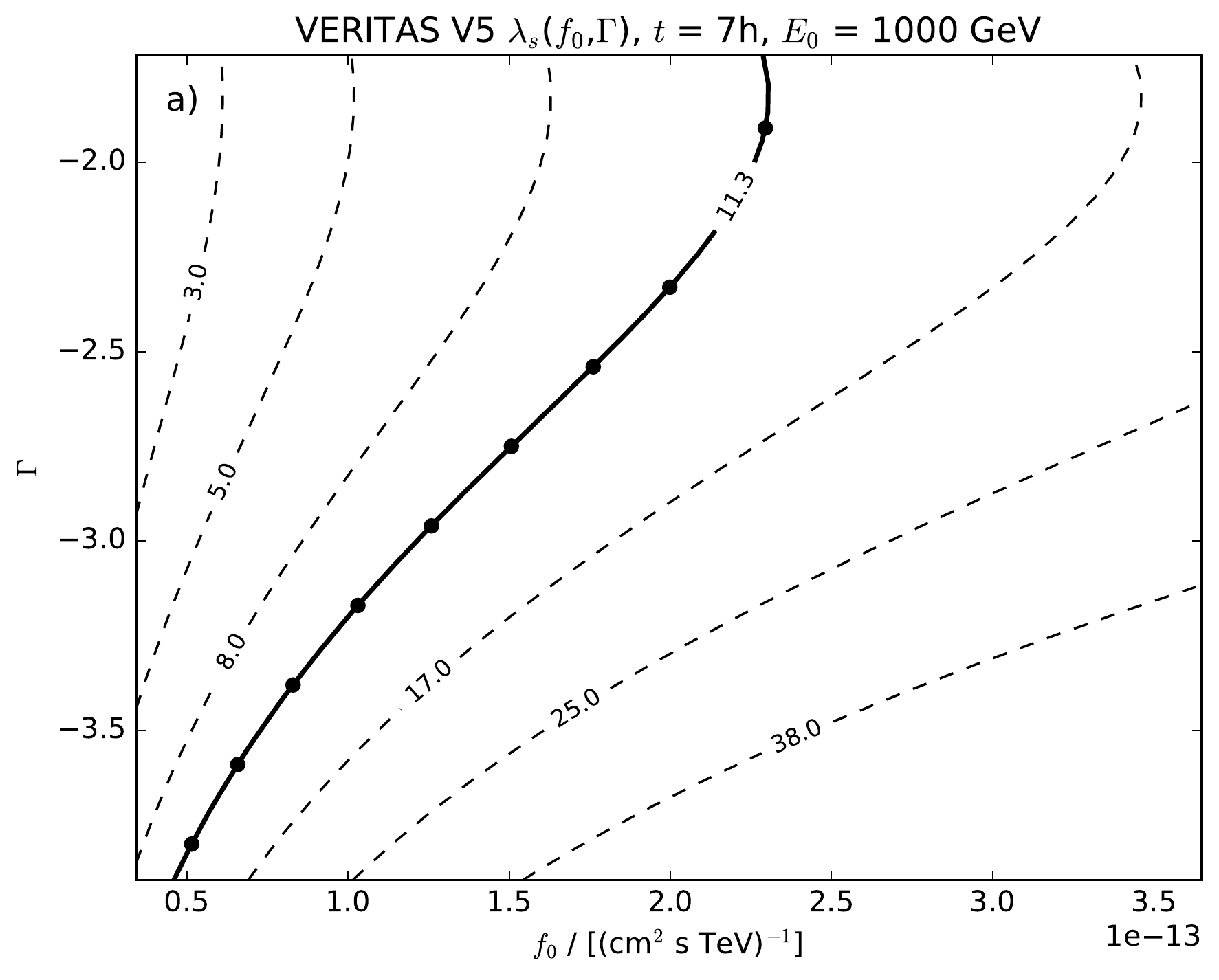}
\includegraphics[width=0.49\textwidth]{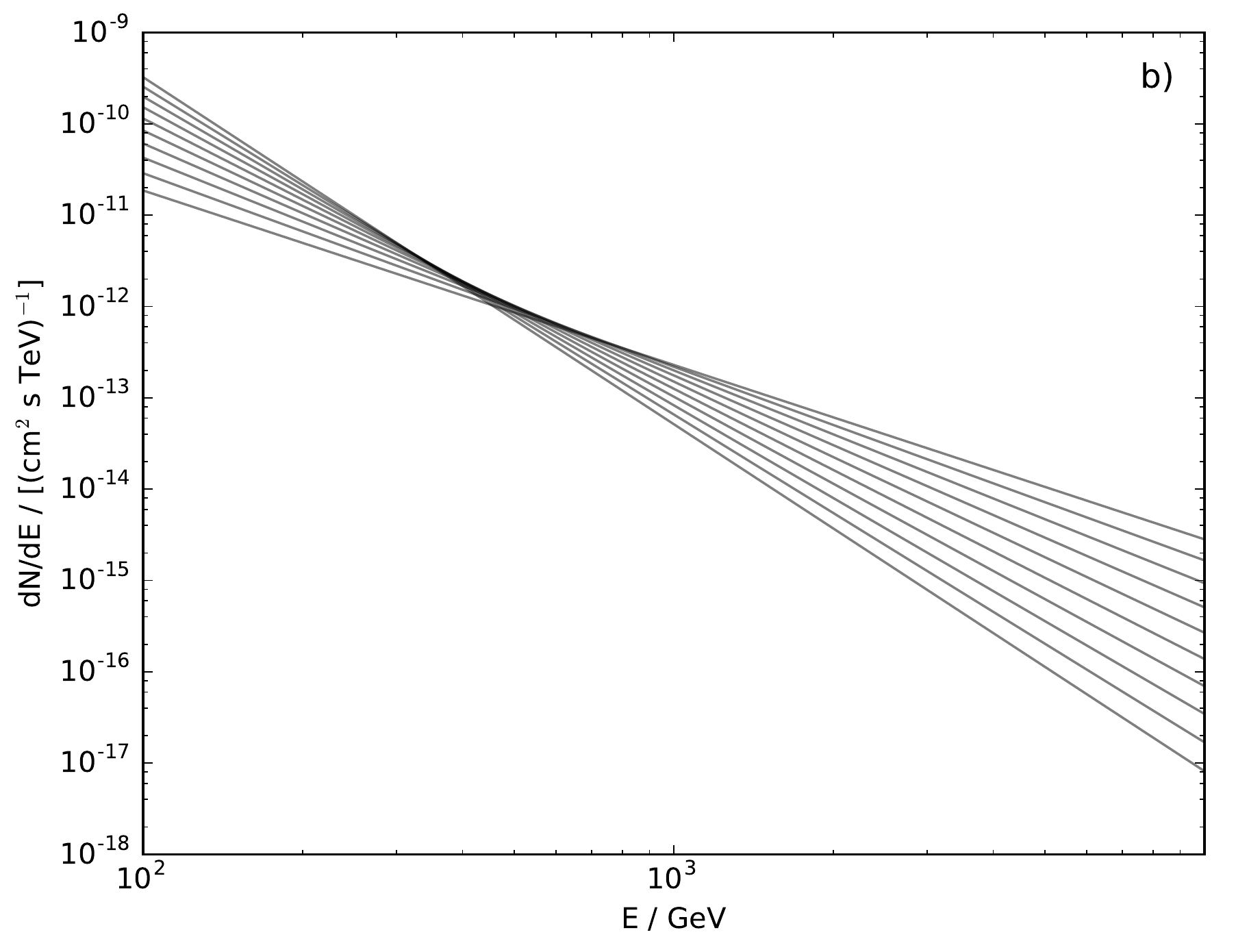}
\caption{
a) The average signal counts as a function of the assumed power law parameters $\lambda_\text{s}(f_0,\Gamma)$ (see Eqn.~\ref{eqn:signal_counts}) for a real observation 
example from the VERITAS Cherenkov telescope~\citep{Mccutcheon2012}.
The solid line
marks $\lambda_\text{lim}$. The power laws 
on this line represent the border of the excluded parameter space (to the 
right of the solid line). Dots indicate example power laws. b)
Example power laws drawn from the
family of curves on the VERITAS M13 exclusion line represented in the spectrum.
It becomes apparent that the family of power laws on the exclusion line
produces an envelope curve when regarding flux upper 
limits in the spectrum parameter space.
}
\label{fig:parameter_space_representation}
\end{figure*}
In the following I derive the novel integral upper limit method, 
and compare it to,
a data set from the VERITAS Cherenkov telescope~\citep{Mccutcheon2012}.
In it, upper limits from an observation of the globular cluster 
M13 are reported. M13 is 
measured to host millisecond pulsars~\citep{Hessels2007} and is 
a popular, yet undetected, target for Cherenkov telescopes
~\citep{Mccutcheon2012,Anderhub2009}. 
In this case the observations took place during the VERITAS
epoch V5~\citep{Park2015} in 2010.
The following components must be specified to calculate 
any integral upper limit
for a specific On/Off measurement:
\begin{enumerate}
  \item $A_\text{eff}(E)$: Effective area as a function of true energy after all cuts
  \item $t$: Effective observation time 
  \item $\lambda_\text{lim}$: Choice of criterion on the 
  signal counts limit e.g.,~\cite{Rolke2005,Knoetig2014}; calculation relies on 
  \begin{enumerate}
    \item Choice of credible interval or confidence interval e.g., $95\%, 99\%$
    \item $\alpha$: On/Off region exposure ratio
    \item $N_\text{on}$: Number of events in the On region
    \item $N_\text{off}$: Number of events in the Off region
  \end{enumerate}
\end{enumerate}
The components used in this section are shown in Tab.~\ref{tab:veritas_m13_obs}.
The resulting signal limit $\lambda_\text{lim}$ 
is calculated using the 95\% credible limit 
on the signal posterior using the method from~\citet{Knoetig2014}.

When investigating spectra of sources it is important to 
take the difference between estimated energy $E_\text{est}$ and 
true (or simulated true) energy $E$ into account. 
This relationship is the migration matrix. 
This means that the effective area measured over 
estimated energy $A_\text{eff}(E_\text{est})$ is a convolution
of the effective area over true energy $A_\text{eff}(E)$,
the migration matrix, and the source spectrum. 
However, as the method proposed here is an integral limit method, 
it needs the effective area as a function of the true (simulated) energy 
$A_\text{eff}(E)$.
In my case however, I use the commonly reported $A_\text{eff}(E_\text{est})$
~\citep{Mccutcheon2012,Aleksic2016Performance}
as an approximation to $A_\text{eff}(E)$ 
as no suitable effective areas are published together with the necessary data
components.
\begin{deluxetable}{ccccccc}

%% Keep a portrait orientation

%% Over-ride the default font size
%% Use Default (12pt)
\tablewidth{11pt}
%% Use \tablewidth{?pt} to over-ride the default table width.
%% If you are unhappy with the default look at the end of the
%% *.log file to see what the default was set at before adjusting
%% this value.

%% This is the title of the table.
\tablecaption{VERITAS M13 observations and $\lambda_\text{lim}$ calculation \label{tab:veritas_m13_obs}}
%% This command over-rides LaTeX's natural table count
%% and replaces it with this number.  LaTeX will increment 
%% all other tables after this table based on this number
\tablenum{1}
%% The \tablehead gives provides the column headers.  It
%% is currently set up so that the column labels are on the
%% top line and the units surrounded by ()s are in the 
%% bottom line.  You may add more header information by writing
%% another line between these lines. For each column that requries
%% extra information be sure to include a \colhead{text} command
%% and remember to end any extra lines with \\ and include the 
%% correct number of &s.
\tablehead{\colhead{$A_\text{eff}$} & \colhead{$t$} & \colhead{C.I.} & \colhead{$\alpha$} & \colhead{$N_\text{on}$} & \colhead{$N_\text{off}$} & \colhead{\boldmath{$\lambda_\text{lim}$}} \\ 
\colhead{} & \colhead{(h)} & \colhead{} & \colhead{} & \colhead{} & \colhead{} & \colhead{} } 
%% All data must appear between the \startdata and \enddata commands
\startdata
1) & 7 & 0.95 & 1/10 & 55 & 670 & \textbf{11.3} \\
\enddata
%% Include any \tablenotetext{key}{text}, \tablerefs{ref list},
%% or \tablecomments{text} between the \enddata and 
%% \end{deluxetable} commands

%% General table comment marker
\tablecomments{The On/Off measurement parameters from Tab.~5.2 and Fig.~5.10 
of~\citet{Mccutcheon2012}. The $\lambda_\text{lim}$ critereon chosen in this paper is
from~\citet{Knoetig2014}
using a 95\% credible interval. The calculated $\lambda_\text{lim}$ is shown in bold.}
%% General table references marker
\tablerefs{1)~\citet{Mccutcheon2012}}
\end{deluxetable}

When examining the family of excluded power laws 
in the $\{f_0,\Gamma\}$ parameter space 
for the VERITAS M13 observation, 
the result is Fig~\ref{fig:parameter_space_representation}a). It shows 
the average signal counts as a function of the power law parameters
$\lambda_\text{s}(f_0,\Gamma)$ (Eqn.~\ref{eqn:signal_counts}),
together with the implicitly defined curve where the 
average signal counts $\lambda_\text{s}$ 
are equal to the signal limit $\lambda_\text{lim}$ (Eqn.~\ref{eqn:exclusion_curve}).
The power laws on this curve represent the border of the 
excluded parameter space ---
the region with higher fluxes to the right is excluded. 
This plot can already be used to compare models with limits, as sources 
populate the power law parameter space~\citep{Ackermann2016FHL}. 
However, most physicists compare source spectra and models in 
the spectrum parameter space.

The following translation of the implicitly defined power law exclusion curve
(Fig.~\ref{fig:parameter_space_representation}a)
into the spectrum parameter space (Fig.~\ref{fig:parameter_space_representation}b)
is key.
When looking at a set of power laws from the 
family of curves on the exclusion line,
it becomes apparent that they
produce an envelope curve when regarding flux upper 
limits in the spectrum parameter space. This means that
for every assumed power law
index $\Gamma$ there is an energy I call \textit{sensitive energy} 
$E_\text{sens}(\Gamma)$ in the 
spectrum parameter space. At the sensitive energy $E_\text{sens}(\Gamma)$ 
the limit power law with 
$\Gamma$ has, locally, the maximum
flux $dN/dE$ (Eqn.~\ref{eqn:power_law}), compared to the other power laws
on the power law exclusion curve. 
Therefore, one specific integral flux upper limit is distinguished at 
$E_\text{sens}(\Gamma)$, which can be used to construct a one-to-one 
relation from points on the power law exclusion curve to points in the
spectrum parameter space by
\begin{equation}
\begin{aligned}
\left(\frac{dN}{dE}\right)_\text{lim}(E) & = & \\
& \underset{f_0,\Gamma}{\text{maximize}}
& & \frac{dN}{dE}(E) ,\\
& \text{subject to}
& & \lambda_\text{s} = \lambda_\text{lim}.
\label{eqn:spectral_exclusion}
\end{aligned}
\end{equation}
The curve $\left(dN/dE\right)_\text{lim}(E)$ in 
the spectrum parameter space represents the border of the region, 
which I call \textit{integral spectral exclusion zone}.
\begin{figure*}
\centering
\includegraphics[width=0.49\textwidth]{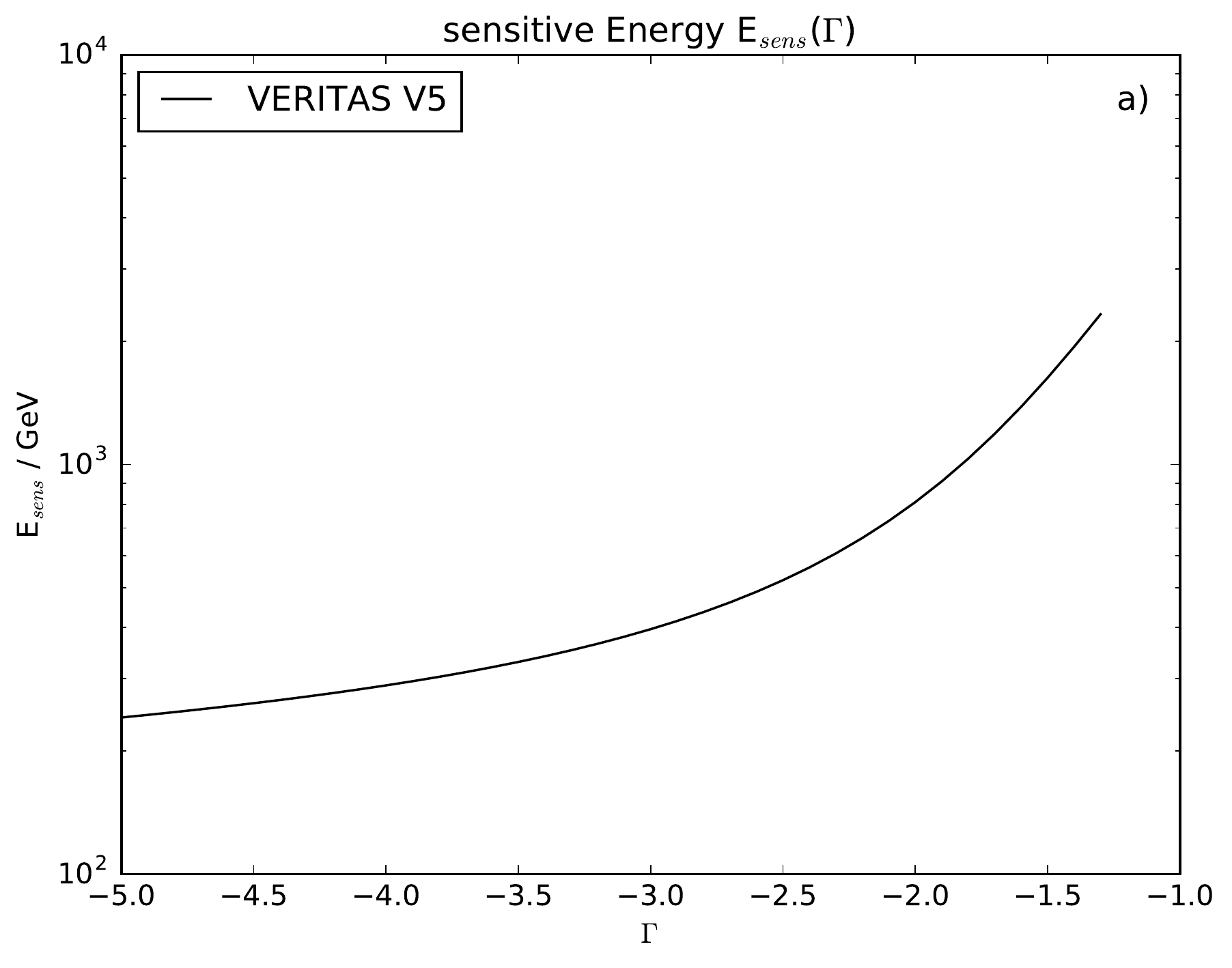}
\includegraphics[width=0.49\textwidth]{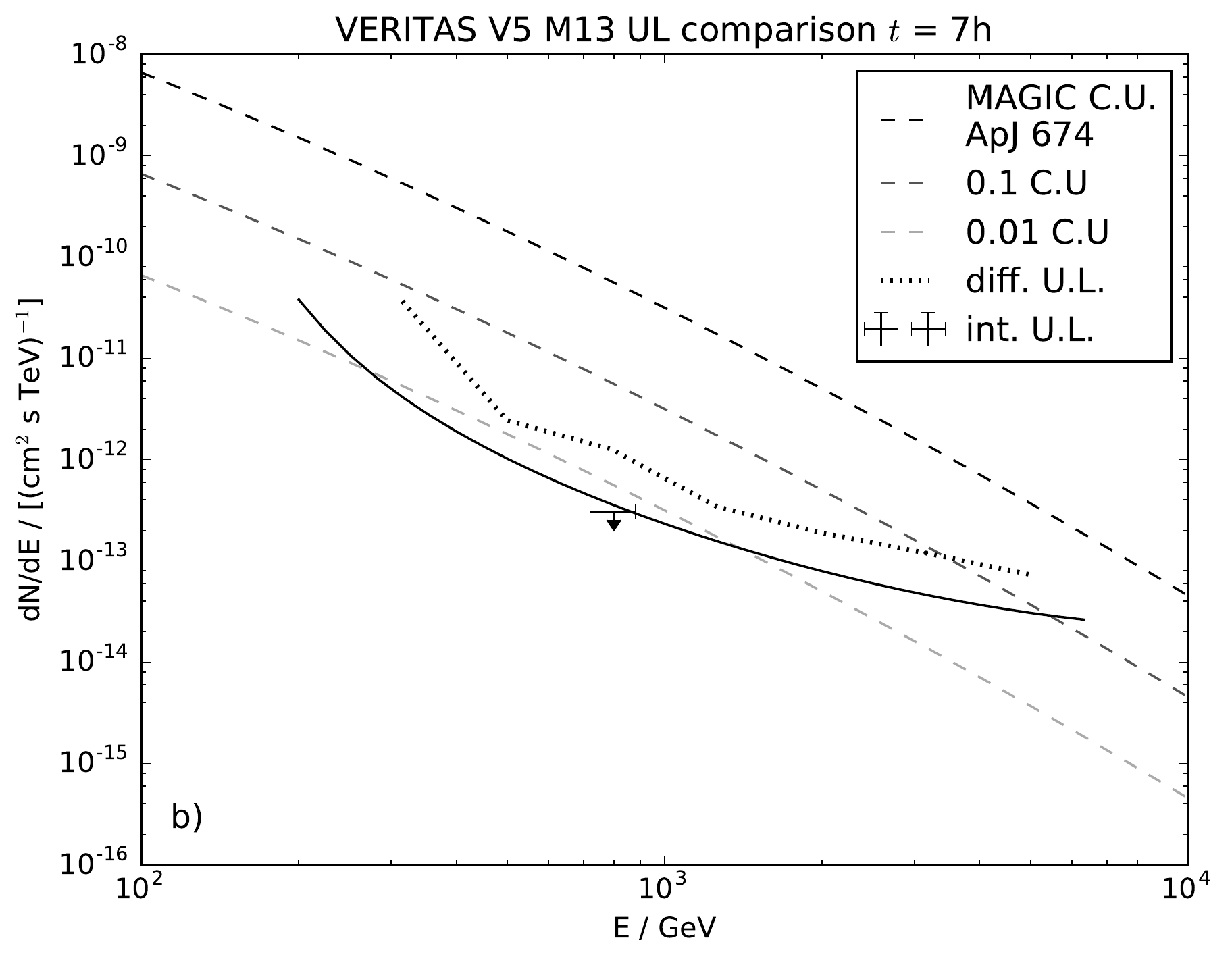}
\caption{a) The \textit{sensitive energy} $E_\text{sens}(\Gamma)$ given the 
effective area of Fig 5.10 in~\citet{Mccutcheon2012}. Harder spectra are
better constrained at higher energies, softer spectra are better 
constrained at lower energies.
b) The \textit{integral spectral exclusion zone} (solid line),
marking the physical boundary above which every power law would have 
been detected, independent of its index $\Gamma$.
It is shown in comparison to the reported limits from~\citet{Mccutcheon2012}.
The reported differential upper limits are about five times weaker 
than the integral spectral exclusion zone.
The VERITAS integral flux limit at the decorrelation energy 
is compatible with the integral spectral exclusion zone at this one point.
Furthermore, the spectrum of the Crab nebula~\citep{Albert2008} is
shown for comparison.}
\label{fig:e_sens_spectral_representation}
\end{figure*}
The necessary conditions for the existence of the integral spectral 
exclusion zone
can be deduced from the method of Lagrange multipliers, 
which is done in Appendix~\ref{app:lagrange}. A fundamental
analytical result of this is 
\begin{eqnarray}
E_\text{sens}(\Gamma) &=& \exp (\mu(\Gamma)),
\label{eqn:e_sens}
\end{eqnarray}
where $\mu(\Gamma)$ is defined as the average of 
the natural logarithm of the true
energy $E$ over sensitive area weighted by the power law
\begin{eqnarray}
\mu(\Gamma) &:=& \frac{\int E^\Gamma A_\text{eff}(E) \ln (E) dE}
{\int E^\Gamma A_\text{eff}(E) dE}.
\label{eqn:mu_mean}
\end{eqnarray}
The example
VERITAS $E_\text{sens}(\Gamma)$ is shown in Fig.
~\ref{fig:e_sens_spectral_representation}a).
To calculate the integral spectral exclusion zone
$\left(dN/dE\right)_\text{lim}(E)$, one has to determine the 
power law $\{f'_0,\Gamma'\}$ that is locally 
tangent to the border of the zone
at energy $E$. This can be done in two ways:
\begin{enumerate}
\item Use the Lagrange multiplier results.
\begin{enumerate}
\item Calculate $\Gamma'$ numerically by inverting $E = E_\text{sens}(\Gamma')$ 
(using Eqn.~\ref{eqn:e_sens}). 
\item Calculate $f'_0$ by solving the power law exclusion line constraint Eqn.~\ref{eqn:exclusion_curve}
\begin{equation}
f_0' = \frac{\lambda_\text{lim}}{t \, c(\Gamma')},
\label{eqn:f_0}
\end{equation}
where $c(\Gamma')$ is defined as the spectrum-weighted acceptance
\begin{equation}
c(\Gamma') := \int \left(\frac{E}{E_0} \right)^{\Gamma'} A(E) dE .
\label{eqn:flux_weighted_acceptance_in_main}
\end{equation}
\item Insert parameters in
\begin{equation}
\left(\frac{dN}{dE}\right)_\text{lim}(E) = 
\frac{dN}{dE}(E)|_{f'_0,\Gamma'}.
\end{equation}
\end{enumerate}

\item Calculate Eqn.~\ref{eqn:spectral_exclusion} 
by maximizing the local flux under constraints directly.
\end{enumerate}

What is the physical meaning of $E_\text{sens}(\Gamma)$ and 
the integral spectral exclusion zone? 
In case a power law with index $\Gamma_\text{cross}$ crosses into the 
integral spectral exclusion zone, its corresponding implicitly defined 
(Eqn.~\ref{eqn:exclusion_curve}) upper limit flux normalization 
$f_0(\Gamma_\text{cross})$ at 
the sensitive energy 
$E_\text{sens}(\Gamma_\text{cross})$ is lower. 
Therefore, the border of the integral spectral exclusion zone
is the physical boundary where every source with a power law spectrum 
crossing it would be detected, independent of $\Gamma_\text{cross}$. 
From this point of view, $E_\text{sens}(\Gamma_\text{cross})$ is the energy at which a 
source with a given power law index $\Gamma_\text{cross}$ is best constrained. 
As the sensitive energy is a monotonically increasing function of 
$\Gamma$ (see Appendix~\ref{app:lagrange}), this yields the intuitive result 
that harder spectra are
better constrained at higher energies and that softer spectra are better 
constrained at lower energies.

Applying this method to the VERITAS M13 data set
results in Fig~\ref{fig:e_sens_spectral_representation}b). 
Here, the integral spectral exclusion zone is calculated
maximizing the logarithm of the flux under constraints directly (see  Eqn.~\ref{eqn:spectral_exclusion}) using
the scipy.optimize Python library~\citep{Jones2001}. The 
integral spectral exclusion zone
is compatible with the reported integral limit at the decorrelation energy
and is about 5 times more sensitive than the reported differential limit.
The VERITAS method of reporting an integral upper limit
at the decorrelation energy (see Sec.~\ref{sec:context}) approximates the 
integral spectral exclusion zone at one point, which is also apparent from 
the method. In this sense the integral spectral exclusion zone
is a generalization of the decorrelation energy method.

\section{Application: The sensitivity of astroparticle telescopes} \label{sec:application}
One important application of the integral spectral exclusion zone
is the fundamental question:
What sources can a certain instrument detect? 
After all, upper limits are a statement of sensitivity. Therefore,
the current methods tackling this question are equivalent to
the current methods of calculating upper limits described in Sec.~\ref{sec:context}. 

First, there are integral methods assuming a certain power law index $\Gamma$.
Among the possible power 
laws, the power law with 
minimum flux normalization $f_0$ which is
required for a detection in a certain observation time is chosen. 
This is often reported as a function of observation time 
~\citep{Aharonian2006Performance}. A similar approach 
is to report the integral 
sensitivity as a function of the minimum flux
required for detection above a given 
threshold energy~\citep{Aleksic2016Performance}. However, this method 
does not constitute a full analysis as the last cut (in energy) is unspecified.
Therefore, it can not be compared to other sensitivities easily.
Furthermore, the power law index
problem persists: What if the source has a different $\Gamma$?

Second, authors report sensitivities as differential upper 
limits~\citep{Aleksic2016Performance,Bernlohr2013},
solving the power law index issue. 
This method is also used to compare instrument sensitivities~\citep{Funk2013Comparison}
but, as discussed in Sec~\ref{sec:context}, being 
a differential limit, it falls short of the real detection capabilities of the instruments.

I demonstrate my novel method using  
the latest published performance data from the MAGIC Cherenkov
telescope~\citep{Aleksic2016Performance}. 
For a given On/Off measurement analysis, the following components 
define the sensitivity of an astroparticle telescope 
to detect sources with power law spectra:
\begin{enumerate}
  \item $A_\text{eff}(E)$: Effective area as a function of true energy after all cuts
  \item $\lambda_\text{lim}$: The sensitivity consensus criterion is, unlike the 
  upper limit criterion, five standard deviations calculated 
  according to~\cite{Li1983}. Subsequently required are
  \begin{enumerate}
    \item $\alpha$: On/Off normalization factor
    \item $\sigma_\text{bg}$: Background event rate in the On region, estimated 
    from simulations or independent Off region data
  \end{enumerate}
\end{enumerate}
Unfortunately, it is not common for Cherenkov telescope collaborations to publish 
$A_\text{eff}(E)$, $\alpha$, and $\sigma_\text{bg}$
coherently for selected and specific analyses. This means that it
is not possible, outside of 
a telescope collaboration, to 
calculate the time until an analysis detects a power law source without 
further assumptions. 
I therefore make the following assumptions (see Tab~\ref{tab:magic_performance})
in order to use the published MAGIC results~\citep{Aleksic2016Performance}:
\begin{enumerate}
  \item $\alpha = 1/5$: Possible values for $\alpha$ stated in that publication
  are $1/3$ and $1/5$. Given that similar Cherenkov 
  telescopes use even more Off regions (see Tab.~\ref{tab:veritas_m13_obs}),
  I assume the more sensitive value $1/5$ is reasonable.%, in particular 
  %when regarding the second additional assumption, which will restrict the analysis to 
  %high-energy events.

  \item $A_\text{eff}(E)$:
  The reported effective area is not given after 
  all cuts. The authors used an additional
  energy cut at $\sim \unit[300]{GeV}$ to achieve the 
  most sensitive instrument. Therefore,
  to approximate $A_\text{eff}(E)$, I use the reported $A_\text{eff}(E_\text{est})$
  with a threshold at $\unit[300]{GeV}$ convoluted with the energy resolution
  stated in~\citet{Aleksic2016Performance}.
\end{enumerate}

Table~\ref{tab:magic_performance} further contains the MAGIC analyses specific zenith distance domain names
(low $Zd \in [0^{\circ},30^{\circ}]$ and medium $Zd \in [30^{\circ},45^{\circ}]$) and 
background rates $\sigma_\text{bg}$.
%% The values (usually only l,r and c) in the last part of
%% \begin{deluxetable}{} command tell LaTeX how many columns
%% there are and how to align them.
\begin{deluxetable}{ccccc}

%% Keep a portrait orientation

%% Over-ride the default font size
%% Use Default (12pt)

%% Use \tablewidth{?pt} to over-ride the default table width.
%% If you are unhappy with the default look at the end of the
%% *.log file to see what the default was set at before adjusting
%% this value.

%% This is the title of the table.
\tablecaption{Inputs for sensitivity calculations of the MAGIC Cherenkov
telescope \label{tab:magic_performance}}

%% This command over-rides LaTeX's natural table count
%% and replaces it with this number.  LaTeX will increment 
%% all other tables after this table based on this number
\tablenum{2}

%% The \tablehead gives provides the column headers.  It
%% is currently set up so that the column labels are on the
%% top line and the units surrounded by ()s are in the 
%% bottom line.  You may add more header information by writing
%% another line between these lines. For each column that requries
%% extra information be sure to include a \colhead{text} command
%% and remember to end any extra lines with \\ and include the 
%% correct number of &s.
\tablehead{\colhead{mode} & \colhead{$A_\text{eff}$} & \colhead{$\alpha$} & \colhead{$\sigma_\text{bg}$} & \colhead{extra cut} \\ 
\colhead{} & \colhead{} & \colhead{} & \colhead{(1/h)} & \colhead{} } 

%% All data must appear between the \startdata and \enddata commands
\startdata
low Zd & 1) & 1/5 & 7.37 & $E_\text{est}>\unit[300]{GeV}$ \\
med. Zd & 1) & 1/5 & 28.83 & $E_\text{est}>\unit[300]{GeV}$ \\
\enddata

%% Include any \tablenotetext{key}{text}, \tablerefs{ref list},
%% or \tablecomments{text} between the \enddata and 
%% \end{deluxetable} commands

%% General table comment marker
\tablecomments{MAGIC 
published low zenith distance parameters ($Zd \in [0^{\circ},30^{\circ}]$) 
and medium zenith distance parameters ($Zd \in [30^{\circ},45^{\circ}]$). 
The On/Off normalization 
factor is assumed to be $\alpha = 0.2$ in this work. 
~\cite{Aleksic2016Performance} report further cutting in energy
to achieve the most sensitive instrument. 
Therefore, I also perform an additional cut in energy, which translates into a cutoff 
in the effective area and a reduced background rate.
}

%% General table references marker
\tablerefs{1)~\cite{Aleksic2016Performance}}

\end{deluxetable}

The integral spectral exclusion zone
method can now be applied to the MAGIC sensitivity 
question. When investigating the sensitivity,
most authors~\citep{Aleksic2016Performance,Aharonian2006Performance,Park2015} 
use five standard deviations calculated 
according to Equation 17 in~\citet{Li1983} as sensitivity
limit criterion.
The connection between this criterion and 
the signal counts limit $\lambda_\text{lim}$
can be constructed from
assuming a constant background flux $\sigma_\text{bg}$ 
in an area with On region exposure
\begin{eqnarray}
N_\text{on} &=& ( \sigma_\text{lim} + \sigma_\text{bg} ) \, t 
\label{eqn:non_over_t} ,\\
N_\text{off} &=& \sigma_\text{bg} \, t / \alpha 
\label{eqn:noff_over_t},
\end{eqnarray}
where the signal rate limit $\sigma_\text{lim}$ (to be inferred) is related to the 
signal counts limit via the observation time
\begin{equation}
\lambda_\text{lim} = \sigma_\text{lim} \, t \label{eqn:lambda_lim_lima}.
\end{equation}
When inserting Eqn~\ref{eqn:non_over_t} and Eqn~\ref{eqn:noff_over_t}
into the limit condition $S_\text{LiMa}(N_\text{on},N_\text{off},\alpha)=5$, one can 
numerically solve for $\sigma_\text{lim}$. 
Then, Eqn.~\ref{eqn:lambda_lim_lima} can be used to calculate 
$\lambda_\text{lim}$.
The Li\&Ma criterion is only valid for count numbers 
which are ``not too few"~\citep{Li1983}. Therefore, it is 
common~\citep{Aleksic2016Performance,Park2015,Aharonian2006Performance} to use
an additional limit, 
which is imposing to measure 
at least ten gamma-ray 
signal counts in the On region $\lambda_\text{lim} \geq 10$.

In this way, the instrument- and analysis-specific 
sensitivities can be calculated 
for any observation time $t$ using the 
integral spectral exclusion zone method.
\begin{figure}
\centering
\includegraphics[width=0.49\textwidth]{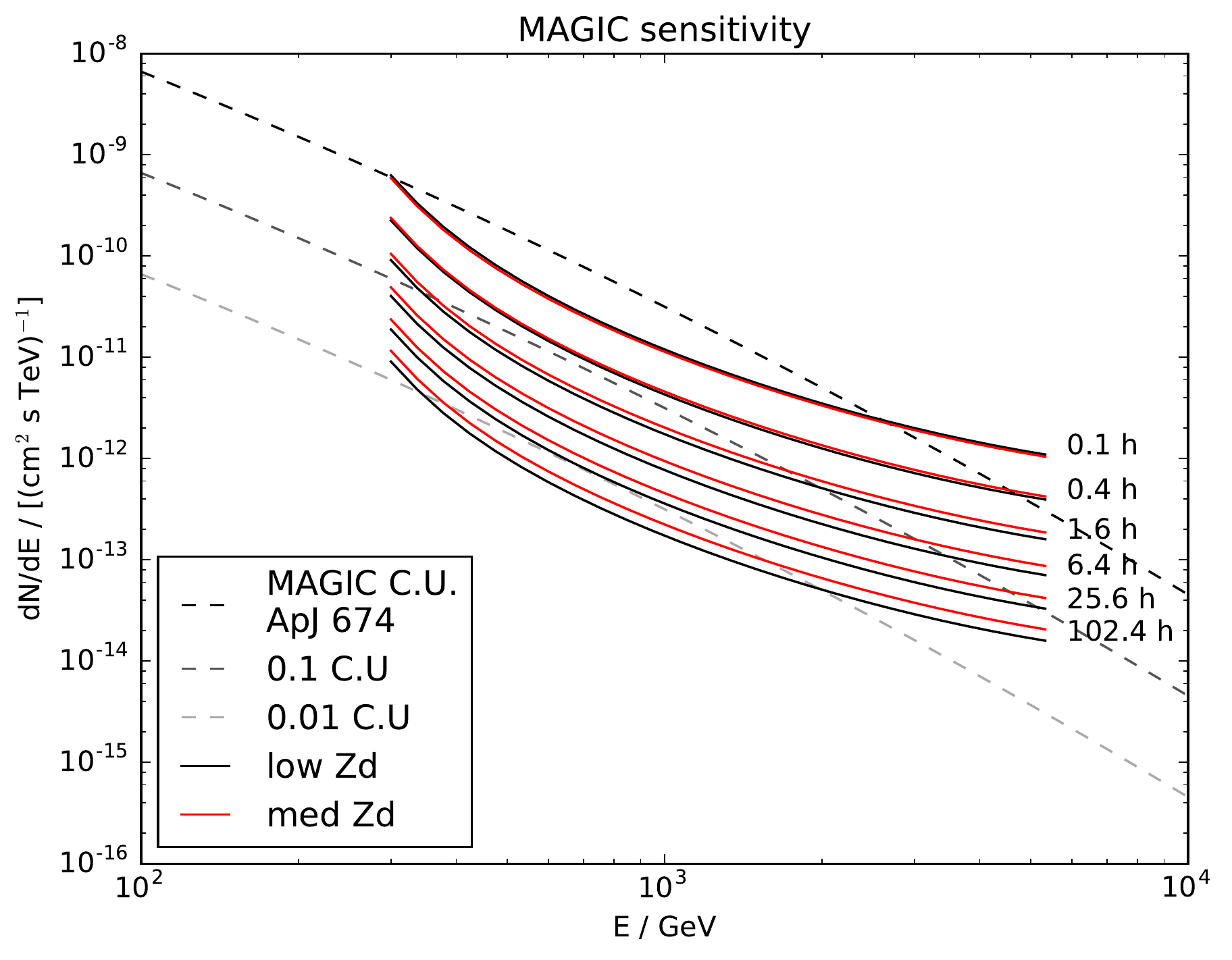}
\caption{The sensitivity of the MAGIC telescope to detect power law sources
for the analysis parameters from Tab~\ref{tab:magic_performance},
independent of $\Gamma$.
It is calculated using the
integral spectral exclusion zone method, for selected observation 
times $t$ and the zenith distance ranges 
low $Zd \in [0^{\circ},30^{\circ}]$ and medium $Zd \in [30^{\circ},45^{\circ}]$ 
as specified in 
\citet{Aleksic2016Performance}. 
Crab nebula~\citep{Albert2008} like spectra (Crab unit C.U.) are shown
for comparison.}
\label{fig:magic_sensitivity}
\end{figure}
%% The values (usually only l,r and c) in the last part of
%% \begin{deluxetable}{} command tell LaTeX how many columns
%% there are and how to align them.
\begin{deluxetable*}{cccccccccccccccc}

%% Keep a portrait orientation

%% Over-ride the default font size
%% Use Default (12pt)

%% Use \tablewidth{?pt} to over-ride the default table width.
%% If you are unhappy with the default look at the end of the
%% *.log file to see what the default was set at before adjusting
%% this value.

%% This is the title of the table.
\tablecaption{Table of selected northern hemisphere galactic TeV sources
\label{tab:sources}}

%% This command over-rides LaTeX's natural table count
%% and replaces it with this number.  LaTeX will increment 
%% all other tables after this table based on this number
\tablenum{3}

%% The \tablehead gives provides the column headers.  It
%% is currently set up so that the column labels are on the
%% top line and the units surrounded by ()s are in the 
%% bottom line.  You may add more header information by writing
%% another line between these lines. For each column that requries
%% extra information be sure to include a \colhead{text} command
%% and remember to end any extra lines with \\ and include the 
%% correct number of &s.
\tablehead{\colhead{name} & \colhead{assoc. tevcat} & \colhead{RA} & \colhead{dec} & \colhead{Zd range} & \colhead{$f_0$} & \colhead{$\Gamma$} & \colhead{ref.} & \colhead{\boldmath{$t_\text{est}$}} \\ 
\colhead{} & \colhead{} & \colhead{($^{\circ}$)} & \colhead{($^{\circ}$)} & \colhead{} & \colhead{$(\unit[]{cm^2\,s\,TeV})^{-1}$} & \colhead{} & \colhead{}& \colhead{(h)} } 

%% All data must appear between the \startdata and \enddata commands
\startdata
Crab & TeV J0534+220 & 83.629 & 22.022 & low & $(2.83\pm0.64)\cdot10^{-11}$& $-2.62\pm0.07$ & 1) & \boldmath{$0.036^{+0.010}_{-0.007}$} \\
HESS J1837-069 & TeV J1837-069 & 279.410 & -6.950 & medium & $(5.0\pm0.3)\cdot10^{-12}$ & $-2.27\pm0.06$ & 2) & \boldmath{$0.336^{+0.035}_{-0.031}$}$^*$ \\
W 41 & TeV J1834-087 & 278.690 & -8.780 & medium & $(3.7\pm0.6)\cdot10^{-12}$ & $-2.5\pm0.2$ & 3) & \boldmath{$0.489^{+0.171}_{-0.124}$} \\
Cas A & TeV J2323+588 & 350.806 & 58.808 & medium & $(1.45\pm0.11)\cdot10^{-12}$ & $-2.75\pm0.3$ & 4) & \boldmath{$2.04^{+0.61}_{-0.51}$} \\
LS 5039 & TeV J1826-148 & 276.563 & -14.842 & medium & $(9.1\pm0.7)\cdot10^{-13}$ & $-2.53\pm0.07$ & 5) & \boldmath{$5.69^{+0.90}_{-0.72}$}$^*$ \\
W 49B & TeV J1911+090 & 287.780 & 9.087 & low & $(3.2\pm1.1)\cdot10^{-13}$ & $-3.14\pm0.34$ & 6) & \boldmath{$16^{+16}_{-7}$}$^*$\\
CTB 87 & TeV J2016+372 & 304.008 & 37.220 & low & $(3.1\pm1.5)\cdot10^{-13}$ & $-2.3\pm0.5$ & 7) & \boldmath{$28^{+58}_{-16}$}$^*$ \\
3C 58 & TeV J0209+648 & 31.379 & 64.841 & medium & $(2.0\pm1.0)\cdot10^{-13}$ & $-2.4\pm0.4$ & 8) & \boldmath{$100^{+230}_{-60}$} \\
\enddata

%% Include any \tablenotetext{key}{text}, \tablerefs{ref list},
%% or \tablecomments{text} between the \enddata and 
%% \end{deluxetable} commands

%% General table comment marker
\tablecomments{Table showing selected galactic TeV gamma-ray sources and
their estimated observation time until detection $t_\text{est}$.\\
Right ascension and declination are given for the epoch J2000.
Calculated values are given in bold. Stars indicate\\predictions, as these sources have not been 
reported by MAGIC.}

%% General table references marker
\tablerefs{1)~\citet{Aharonian2004}, 2)~\citet{Aharonian2006},
3)~\citet{Albert2006}, 4)~\citet{Kumar2015},\\5)~\citet{Aharonian2006binary} low state,
6)~\citet{Abdalla2016}, 7)~\citet{Aliu2014spatially}, 8)~\citet{Aleksic2015}}

\end{deluxetable*}
\begin{figure*}
\centering
\includegraphics[width=0.49\textwidth]{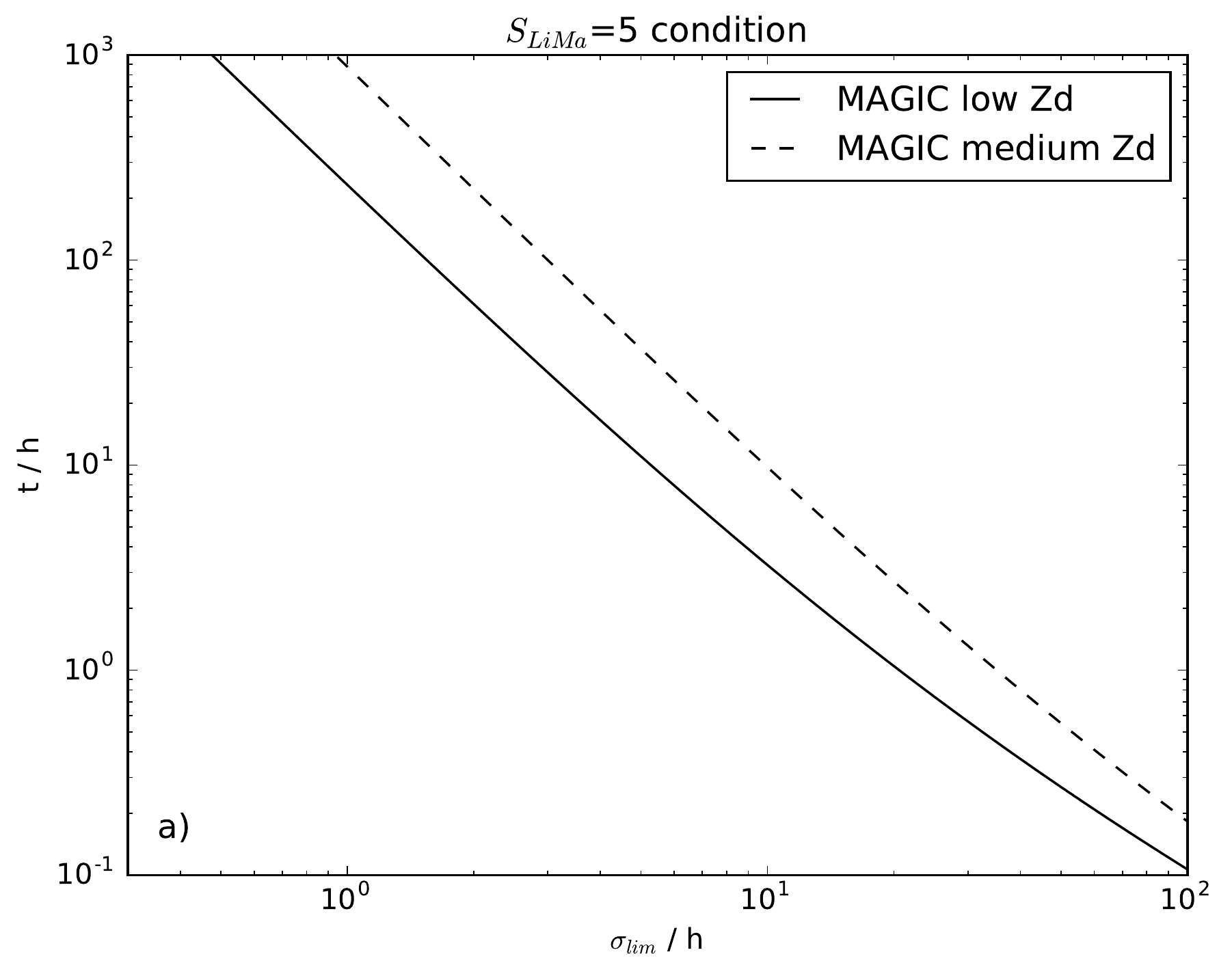}
\includegraphics[width=0.49\textwidth]{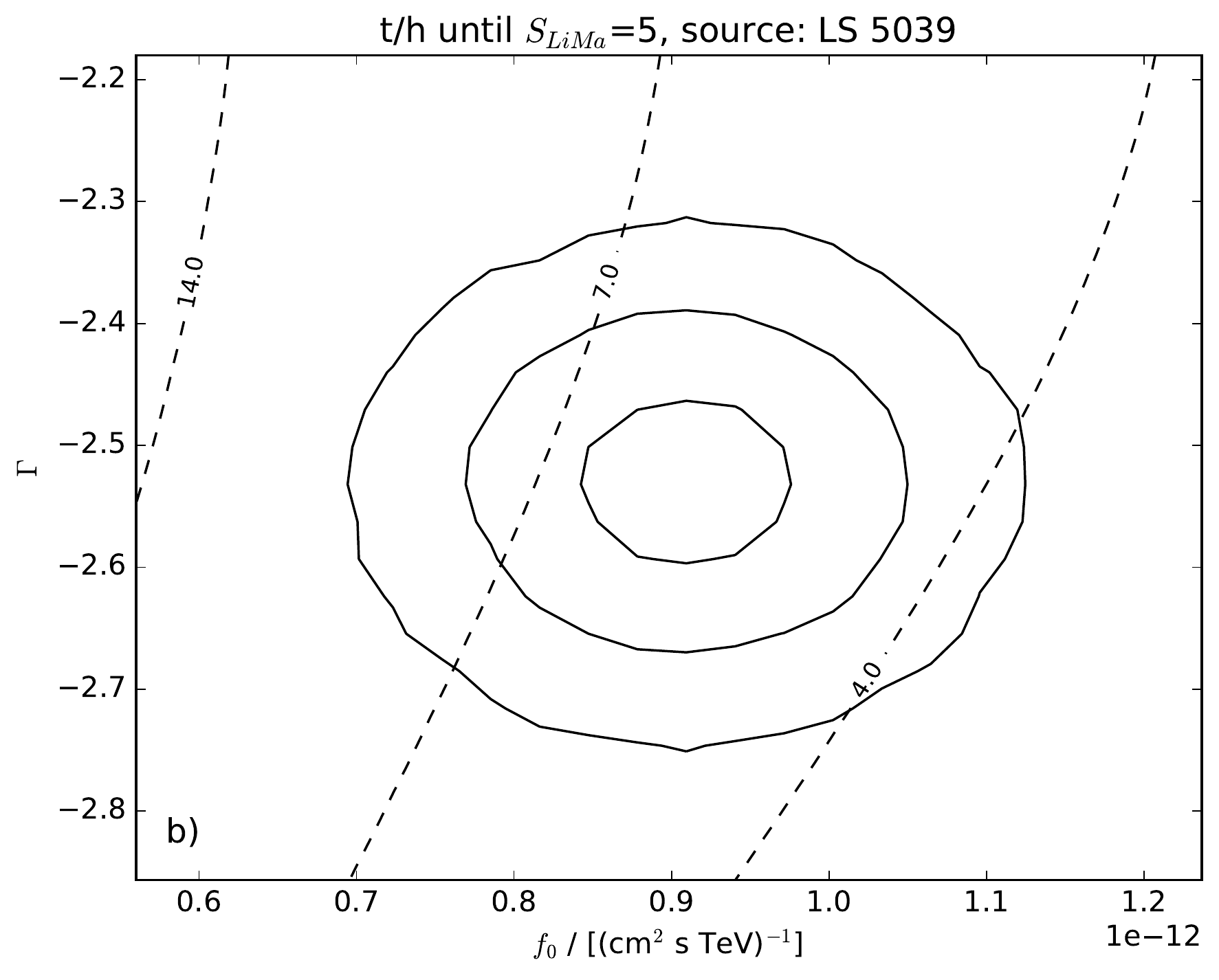}
\caption{a) The average observation time $t$ until a certain signal rate limit 
$\sigma_\text{lim}$ is reached, demonstrated for the MAGIC analyses
shown in Tab~\ref{tab:magic_performance}. b) The average observation 
time $t$ until MAGIC detects a medium Zd source as a function of the power law parameters.
This is shown in comparison to the measured 
LS 5039 power law parameter $1\sigma$, $2\sigma$, and $3\sigma$ confidence levels
~\cite{Aharonian2006binary}.}
\label{fig:sigma_inv_t_ls_5039}
\end{figure*}

The result, applied to the MAGIC telescope
analyses from Tab.~\ref{tab:magic_performance}, is shown in 
Fig.~\ref{fig:magic_sensitivity}. 
It agrees well with MAGIC's ability~\citep{Aleksic2016Performance} 
to detect a Crab nebula-like source with 
1\% Crab nebula flux within $\sim\unit[25]{h}$.

In the case of specific sources, it is useful to 
return to the power law parameter space $\{f_0,\Gamma\}$.
Table~\ref{tab:sources} shows eight selected galactic TeV emitting gamma-ray sources,
observable from the MAGIC site (Latitude $\unit[28.76^{\circ}]{N}$) 
at low or medium zenith distances. This implies source declinations 
$\in [-16^{\circ}.24, 73^{\circ}.76]$. 
As the MAGIC background rate parameters $\sigma_\text{bg}$ 
from Tab.~\ref{tab:magic_performance} are 
valid for point-like sources,  
only MAGIC quasi point-like sources (up to a mean extension of 
$\bar{\rho} < 0.1^{\circ}$) are considered. In the following, I 
use the LS 5039 low state flux as an example for how to 
estimate the necessary observation time $t_\text{est}$ until detection. 

The average time to detect a certain power law in the power law parameter space
must be calculated by numerically inverting  
the signal rate limit $\sigma_\text{lim}(t)$. 
Figure~\ref{fig:sigma_inv_t_ls_5039}a) shows 
$\sigma^{-1}_\text{lim}$ as a function of the signal rate 
for the low Zd and medium Zd MAGIC analyses.

Then, the power law limit curve Eqn~\ref{eqn:exclusion_curve} 
can be reformulated using the
average signal counts function $\lambda_\text{s}$ (Eqn.~\ref{eqn:signal_counts}),
the power law spectrum $dN/dE$ (Eqn.~\ref{eqn:power_law}), and $\sigma^{-1}_\text{lim}$ as
\begin{equation}
t = \sigma^{-1}_\text{lim}(c(\Gamma)\,f_0).
\label{eqn:time_to_detection}
\end{equation}
Finally, the average estimated
observation time $t_\text{est}$ until detection (Eqn.~\ref{eqn:time_to_detection}) 
can be calculated as a function of the power law 
parameters $\{f_0,\Gamma\}$. This is demonstrated in Fig.~\ref{fig:sigma_inv_t_ls_5039}b)
together with the measured LS 5039 parameter confidence intervals~\citep{Aharonian2006binary}
assuming uncorrelated normal distributed errors in $f_0$ and $\Gamma$.
The data indicates that MAGIC, 
using the analysis parameters from Tab.~\ref{tab:magic_performance},
would need $t_\text{est} \sim \unit[4]{h}-\unit[7]{h}$ 
in order to detect this source. 

\begin{figure}
\centering
\includegraphics[width=0.49\textwidth]{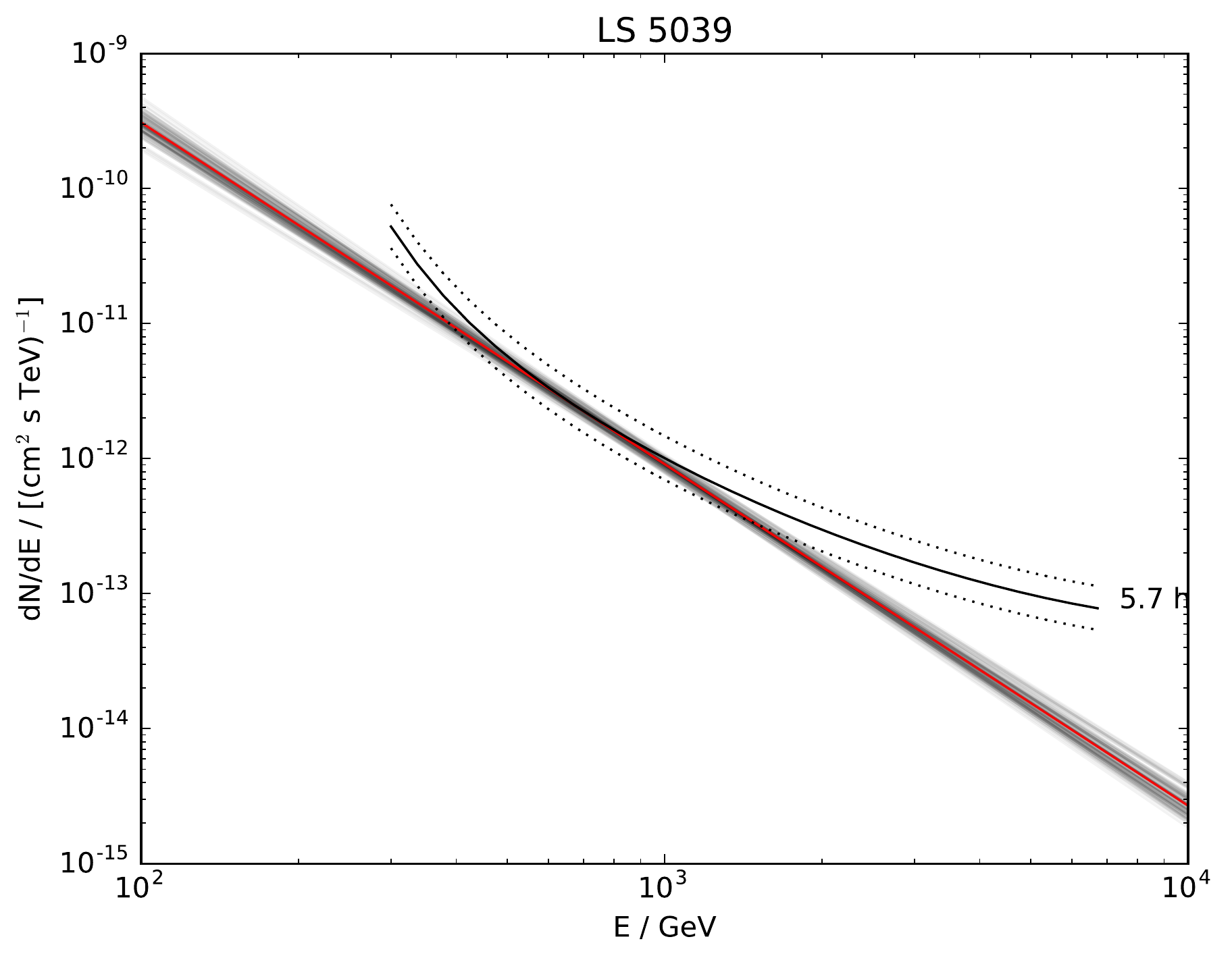}
\caption{Spectra drawn from the parameter space according to the 
LS 5039 parameter errors with a red line indicating the most likely 
power law. Integral spectral exclusion zones are 
calculated for three MAGIC times of observation. The solid 
line indicates the integral spectral exclusion zone for the
median estimated time to detection $t_\text{est} = \unit[5.69]{h}$.
It touches the red most likely spectrum at its sensitive energy. 
The dotted lines indicate the evolution of the integral spectral 
exclusion zone, given half or double the median time to detection.}
\label{fig:integral_spectral_exclusion_zone_ls_5039}
\end{figure}

To quantify $t_\text{est}$, one can
sample from the parameter space according to the 
LS 5039 parameter errors and calculate the 
histogram of times to detection.
I report the times to detection as the median time with asymmetric 
errors calculated from the 16th percentile and the 84th percentile
in Tab~\ref{tab:sources}.
In the case of LS 5039, this results in 
an estimated MAGIC time to detection of $t_\text{est} = \unit[5.69^{+0.90}_{-0.72}]{h}$.
Four out of eight $t_\text{est}$ calculated are predictions
as MAGIC has not reported detections on HESS J1837-069, 
LS 5039, W 49B, and CTB 87.

Finally, one can illustrate the
estimated time to detection
using the integral spectral exclusion zone, which is done in
Fig.~\ref{fig:integral_spectral_exclusion_zone_ls_5039}.
It shows the simple but powerful interpretation of the 
integral spectral exclusion zone as the instrument detection 
capability --- the border of the 
region in the spectrum parameter space which any detectable 
power law would cross.

\section{Discussion}
\label{sec:discussion}
The calculations in this manuscript are done 
using the true energy $E$, as I only consider integral results (see Sec.~\ref{sec:method}). 
Therefore, limited energy resolution and migration
do not affect the integral spectral exclusion zone, 
as a power law source does indeed emit 
with its true individual spectrum and an instrument measures the events according to
$A_\text{eff}(E)$. 
This is the case even when one is unable to reconstruct the individual event energies
at all. Therefore, it is 
possible to know the sensitivity of an instrument to detect power law sources without
reconstructing the energy of individual events.

There are several cases when sources reveal additional curvature, besides 
the general power law behavior, after measuring them with high 
significance and low statistical uncertainty.
However, the upper limit question and the sensitivity question for 
an individual telescope analysis only
consider what happens at the detection threshold
and only what happens in the limited energy domain of $A_\text{eff}(E)$ (see Eqn.~\ref{eqn:signal_counts}).
There, basically every source in astroparticle physics appears as power law
initially. Therefore, I argue that the integral spectral exclusion zone is applicable,
even when source spectra deviate ``not too much'' from power laws.

In this manuscript I only considered On/Off measurements
so far, which infer results in view of Poissonian signal-and-noise
experiments (see Sec.~\ref{sec:context}), but there may be other 
cases when, for instance,
no background is expected and $\lambda_\text{lim}$ can be calculated from 
a single Poisson process producing $N_\text{on}$. In general, there are plenty 
of statistical approaches to choose from. 
However, the details of calculating $\lambda_\text{lim}$
are up to each individual experimentalist
and what method she believes in. The signal count limit
$\lambda_\text{lim}$,
formulated as constraint in Eqn~\ref{eqn:spectral_exclusion},
is only the 
interface of my integral spectral exclusion zone method
with the world of statistics: 
It stays the same while the statistical methods may be exchanged.

\section{Conclusion} \label{sec:conclusion}
Many high-energy astronomers struggle with the implications of 
non-detections and instrument limits. 
The universal observation of power laws in high-energy sources 
allows framing these problems in the parameter space of the power law model.
A simple and powerful one-to-one relation of the 
integral spectral limit in the power law parameter space into spectra
makes it possible to construct the
integral spectral exclusion zone. This 
integral spectral exclusion zone demonstrates 
what astroparticle telescopes are actually capable of detecting 
-- and how long it will take.
In this way
one can easily compare instrument performances, optimize 
analysis algorithms, and test model predictions
of high-energy astrophysics.

A python package implementing the methods from this manuscript 
can be found at \url{https://github.com/mahnen/gamma_limits_sensitivity}

\acknowledgments

I would like to thank Adrian Biland, Sebastian A. Mueller, Scott Lindner, 
Camila Shirota and Dariusz Gora for support, discussions, and valuable comments on the 
article.

\appendix

\section{Application of the Lagrange multiplier method for calculating 
the integral spectral exclusion zone}
\label{app:lagrange}
Motivated by the results shown in 
Fig.~\ref{fig:parameter_space_representation}, I assume the 
existence of a \textit{maximum} flux (Eqn.~\ref{eqn:power_law}) 
given the external constraint (Eqn.~\ref{eqn:exclusion_curve}). 
Then, the necessary conditions for the existence of the integral spectral 
exclusion zone can be deduced from the method of Lagrange multipliers. 
The first step is the construction of the problem specific 
Lagrange function $\mathcal{L}(f_0,\Gamma)$ by joining the function to be maximized $dN/dE$
to the external constraint
using a Lagrange multiplier $\delta$
\begin{equation}
\mathcal{L}(f_0,\Gamma) = f_0 \left(\frac{E}{E_0}\right)^\Gamma + 
\delta \, (t \, f_0 \, c(\Gamma) - \lambda_\text{lim}),
\label{eqn:lagrange_function}
\end{equation}
where $c(\Gamma)$ is an abbreviation for the spectrum weighted acceptance
\begin{equation}
c(\Gamma) := \int \left(\frac{E}{E_0} \right)^\Gamma A(E) dE .
\label{eqn:flux_weighted_acceptance}
\end{equation}
By taking the gradient with respect to the parameters $\{f_0, \Gamma\}$ and 
equating it to zero one 
gets a set of equations
\begin{eqnarray}
f_0 \left(\frac{E}{E_0}\right)^\Gamma \ln\left(\frac{E}{E_0}\right) +
\delta \, t \, f_0 \, \frac{\partial c(\Gamma)}{\partial \Gamma} &= 0 ,
\label{eqn:nec_con_1}
\\
\left(\frac{E}{E_0}\right)^\Gamma +
\delta \, t \, c(\Gamma) &= 0 ,
\label{eqn:nec_con_2}
\end{eqnarray}
which, together with the external constraint Eqn.~\ref{eqn:exclusion_curve},
constitute the necessary conditions.
Equation~\ref{eqn:nec_con_2} solved for $\delta$ gives 
\begin{equation}
\delta = \frac{-\left(\frac{E}{E_0}\right)^\Gamma}{t \, c(\Gamma)}.
\end{equation}
This equation inserted back into Eqn.~\ref{eqn:nec_con_1} results in
\begin{equation}
\ln\left(\frac{E}{E_0}\right) = \frac{1}{c(\Gamma)} 
\frac{\partial c(\Gamma)}{\partial \Gamma} .
\label{eqn:e_sens_step1}
\end{equation}
When interchanging the integral and differentiation in the explicit 
formula of $\partial c(\Gamma) / \partial \Gamma$ and solving Eqn~\ref{eqn:e_sens_step1}
for the energy E the result is
\begin{equation}
E = \exp(\mu(\Gamma)),
\label{eqn:e_sens_appendix}
\end{equation}
where $\mu(\Gamma)$ is an abbreviation for the 
average of the natural logarithm of the
energy $E$ over the spectrum weighted acceptance
\begin{eqnarray}
\mu(\Gamma) &:=& \frac{\int E^\Gamma A_\text{eff}(E) \ln (E) dE}
{\int E^\Gamma A_\text{eff}(E) dE}.
\label{eqn:mu_mean_appendix}
\end{eqnarray}
$\mu(\Gamma)$ is independent of $E_0$, which is a reasonable result for a scale factor.
I identify the energy at which the Lagrange function has a maximum 
with the sensitive Energy $E_\text{sens}$ in Sec~\ref{sec:method}. 

Equation~\ref{eqn:mu_mean_appendix} shows that 
the existence of the integral spectral exclusion zone is
a direct consequence of the physical argument that there can not be infinite counts
which leads to the normalizability of the spectrum weighted acceptance 
(finite numerator and denominator in Eqn.~\ref{eqn:mu_mean_appendix}).
The average natural logarithm of the 
energy $\mu(\Gamma)$ is an increasing function of $\Gamma$, as harder spectra 
produce higher energy events more frequently. Therefore, $E_\text{sens}(\Gamma)$ 
is also an increasing function, which means it is invertible.
Finally, one can take Eqn.~\ref{eqn:e_sens_appendix}, solve it numerically 
for $\Gamma$, and find the last remaining undetermined parameter $f_0$ 
by solving the 
so far unused constraint function Eqn.~\ref{eqn:exclusion_curve}, which is 
shown in Eqn.~\ref{eqn:f_0}.



\bibliography{on_upper_limits_mahnen_aastex}


%% \begin{thebibliography}{}

%% \bibitem[Corrales(2015)]{2015ApJ...805...23C} Corrales, L.\ 2015, \apj, 805, 23
%% \bibitem[Hanisch \& Biemesderfer(1989)]{1989BAAS...21..780H} Hanisch, R.~J., \& Biemesderfer, C.~D.\ 1989, \baas, 21, 780 
%% \bibitem[Lamport(1994)]{lamport94} Lamport, L. 1994, LaTeX: A Document Preparation System, 2nd Edition (Boston, Addison-Wesley Professional)
%% \bibitem[Schwarz et al.(2011)]{2011ApJS..197...31S} Schwarz, G.~J., Ness, J.-U., Osborne, J.~P., et al.\ 2011, \apjs, 197, 31  
%% \bibitem[Vogt et al.(2014)]{2014ApJ...793..127V} Vogt, F.~P.~A., Dopita, M.~A., Kewley, L.~J., et al.\ 2014, \apj, 793, 127  

%% \end{thebibliography}

%% \allauthors

%% \listofchanges

\end{document}